# Probing the strongly correlated magnetic state of $Co_2C$ nanoparticles at low temperatures using μSR

Nirmal Roy[1], P C Mahato[1], Suprotim Saha[1], M. Telling[2], J. S. Lord[2], D T Adroja[2,3], S. S. Banerjee[1,+]

[1]*Indian Institute of Technology Kanpur, Kanpur, Uttar Pradesh 208016, India*

[2]*ISIS Facility, Rutherford Appleton Laboratory, Chilton, Didcot Oxon OX11 0QX, United Kingdom*

[3]*Highly Correlated Matter Research Group, Physics Department, University of Johannesburg, P.O. Box 524, Auckland Park 2006, South Africa*

**Abstract:**

$Co_2C$ nanoparticles (NPs) are amongst transition metal carbides whose magnetic properties have not been well explored. A recent study by Nirmal Roy et al. [1] showed that a collection of $Co_2C$ NPs exhibit exchange bias (EB) effect below $T_{EB}$ = 50 K and also a spin glass (SG) state below $T_{SG}$ = 5 K. We use magnetic, electrical transport, specific heat, and muon spin rotation (μSR) measurements to explore further the magnetic properties of a pellet made with 40 nm diameter pure $Co_2C$ NPs. We uncover the onset of Kondo localization at Kondo temperature $T_K$ (= 40.1 K), which is close to the onset temperature ($T_{EB}$) of the EB effect. A crossover from the Kondo-screened scenario to an RKKY interaction-dominated regime is also observed for $T < T_K$. Temperature-dependent specific heat measurement further supports the Kondo localization scenario in the pellet and shows the heavy fermionic nature of the strongly correlated electronic state in $Co_2C$. The zero field μSR asymmetry spectra in the low-temperature regime are characterized by two distinct fast and slow relaxation rates. The spectra show the absence of long-range magnetic order in the sample. However, our analysis suggests the NPs-pellet shows the presence of a dominant magnetically disordered fraction and a smaller fraction with short-range order. Muons in the disordered fraction exhibit a slower relaxation rate, while muons in the smaller fraction with short-range order exhibit a faster relaxation rate. We observe an increase in this fast relaxation rate between $T_{EB}$ and $T_{SG}$. This increase below $T_{EB}$ ~ 50 K suggests a slowing down of the fluctuating local magnetic environment around muons. Transverse field (TF) - μSR asymmetry spectra show the emergence of a stable, multi-peaked local magnetic field distribution in the pellet below $T_{EB}$. Longitudinal field (LF) μSR spectra shows distinct changes in the dynamics of fluctuations suggesting the presence of a frozen glassy like state below 6 K. Based on our results, we suggest that below $T_{EB}$, the pellet of $Co_2C$ NPs develops a magnetic interface that separates the two magnetic fractions; one is a

**Corresponding authors email**:[+]satyajit@iitk.ac.in

disordered fraction, and the other is a fraction with short-range order. The exchange interaction that sets in below $T_{EB}$ at the interface couples the two fractions, leading to a suppression of the fluctuations. With the suppression of magnetic fluctuations below $T_{EB}$, strong correlation effects in the electronic state of $Co_2C$ lead to Kondo localization.

## I. INTRODUCTION:

In the core-shell nanoparticles (NPs) system, a wide variety of complex effects and phases like the exchange bias (EB) effect, spin glass (SG) phase, spin-liquid phases, the spin ice phase, magnetic proximity effect [2-8], have been reported. The manipulation of core and shell dimensions and exploring different materials produce unique properties in these core-shell [4, 9, 10] NPs. The EB effect relates to two different magnetic species exchange-coupled across an interface. In general, for a ferromagnet (FM) – antiferromagnet (AFM) heterostructure, the system is cooled from above the AFM Néel temperature ($T_N$) to below it by applying a magnetic field that can saturate the FM. The AFM lattice spins in its ordered state via interfacial exchange coupling to the FM spins pins the latter in a preferred direction creating an additional unidirectional anisotropy. This EB interfacial exchange anisotropy results in an asymmetric shift of the magnetization ($M$) vs. the applied field ($H$) hysteresis loops along the field and magnetization axis compared to the typical symmetric loop found in a simple bulk FM. The EB effect is significant for spintronic devices and magnetic information storage technology [11-13]. Therefore, the EB effect continues to be a popular investigated phenomenon studied in core-shell NPs. It was first discovered in the FM cobalt (Co)/AFM cobalt oxide (CoO) system [14]. Since then, EB has been reported in a plethora of systems consisting of clusters of small particles, FM films deposited on a single crystal or polycrystalline AFMs, FM/AFM thin-film bilayers, FM/ferrimagnet, FM/spin-glass heterostructures [15-19]. The fascinating magnetic properties of these NP systems originate from the complex interplay of finite size and surface effects, making them very different from their bulk counterpart [20]. In some magnetic NPs, although the core may be magnetically ordered, there can be local symmetry breaking near the surface of the NPs, producing a magnetically disordered spin configuration structure.

In these NPs, a magnetic core-shell structure naturally emerges with a magnetically ordered core and a disordered spin configuration on the shell. An exchange coupling at the interface between this disordered shell and the ordered core gives rise to EB in these systems [10].

In recent times, bulk magnetization studies of Cobalt Carbide ($Co_2C$) NPs have unraveled some intriguing features of its magnetism at nanoscales. Among the well-known transition metal carbides (TMC), the magnetic properties of $Co_2C$ have not been well explored [21, 22]. TMCs are well known for their extreme hardness, high thermal and electrical conduction, high coercive fields ($H_C$), and high magnetic anisotropy [21-25]. The high magnetic anisotropy makes these TMCs suitable candidates for application as rare-earth-free permanent magnets and as magnetic storage devices if their magnetic properties can be tailored and controlled at the nanoscale [26]. While DFT calculations suggest $Co_2C$ is a paramagnet in the bulk form [27], early studies showed that NPs of $Co_2C$ are ferromagnetic [28, 29]. More recent magnetic measurements [1] show that the ferromagnetism in the sample made from 40 nm diameter $Co_2C$ NPs survives up to room temperature. This result suggested that a high blocking temperature (exceeding 300 K) or large magnetic anisotropy is associated with these NPs. Furthermore, below $T_{EB}$ = 50 K, these NPs exhibit a significant EB effect with an exchange bias field ~ 0.25 kOe at 2 K. Studies also showed that the $Co_2C$ NPs sample exhibits a low-temperature SG phase that appears below $T_{SG}$ ~5 K [1]. Furthermore, DFT investigations [1] suggest the $Co_2C$ may have a structurally ordered core and a structurally disordered shell configuration. These intriguing results suggest a need to probe the local magnetic configuration in the composites made from $Co_2C$ NPs.

Here, using magnetic, electrical transport, thermal and µSR measurements, we explore the internal magnetic properties of a pellet made with 40 nm diameter pure $Co_2C$ NPs. By analysing our resistance (*R*) versus temperature (*T*) behaviour we uncover three distinct regimes for our $Co_2C$ NPs-pellet. The regime III is the high $T$ region > 50 K (which is the exchange bias $T$, $T_{EB}$), where transport is governed by electron-phonon interaction mediated scattering. Below $T_{EB}$, in regime II $R(T)$ behaviour shows Kondo localization and features associated with scattering from these regions. In region I for $T$ < 35 K, the transport behaviour fits to system which not only exhibits Kondo localization but there is interaction between localized sites through Ruderman-Kittel-Kasuya-Yosida (RKKY) interactions. The Kondo localization temperature is estimated to be $T_K$ = 40.1 K. At $T_K$, a broad peak is seen in the magnetic contribution to heat capacity. A large value of the Sommerfeld constant ($\gamma$) suggests $Co_2C$ is a heavy-fermion system with a strongly correlated electronic state. Zero-field (ZF) μSR study shows that the asymmetry spectra are characterized by two distinct relaxation rates, indicating two inequivalent stopping sites in the $Co_2C$ NPs-pellet. We see no evidence of long-range magnetic order in the NPs-pellet in our investigation's temperature range of 1.8 K to 220 K. However, the ZF- μSR spectra analysis suggests the presence of a dominant magnetically disordered fraction with a slower relaxation rate and a smaller fraction with short-range order having a much faster relaxation rate. The rate of the fast-relaxing component is seen to increase between $T_{EB}$ and $T_{SG}$; it decreases below $T_{SG}$. This increase is related to the stabilization of fluctuating magnetic moments below $T_{EB}$ ~ 50 K. Concomitant with the onset of stabilization of the fluctuations, analysis of the transverse field μSR asymmetry spectra (TF- μSR) shows the emergence of a multi-peaked magnetic field distribution inside the NPs below $T_{EB}$. Longitudinal field (LF) μSR spectra shows distinct changes in the dynamics of fluctuations suggesting the presence of a frozen glassy like state below 6 K. Our results suggest that below $T_{EB}$, a magnetic interface appears, which separates two magnetic fractions. One is the fraction with short-range order, and the other is the magnetically disordered fraction. Below $T_{EB}$, the exchange interaction at the interface couples the two magnetic fractions and reduces the fluctuations of the magnetic moments. With the stabilization of fluctuations, the strong correlation effects in Co$_2$C lead to Kondo localization at magnetic ionic sites below $T_{EB}$.

## II. EXPERIMENTAL RESULTS AND DISCUSSION

**Magnetic behavior as a function of *T* of Co$_2$C NP pellet:**

The method for synthesis of 40 nm diameter Co$_2$C NP powders and their structural and morphological characterization using powder XRD and TEM are discussed in supplementary information section I (SI-I) and elsewhere [1, 30]. The powders of Co$_2$C NPs from this batch were compactified into cylindrical pellets with a diameter of 8 mm and thickness of 1 mm, and these pellets were used for magnetic, transport, and thermal measurements. A cylindrical pellet of size 16 mm diameter and thickness 1 mm were also made from the Co$_2$C NPs for the μSR measurements. The magnetic measurements were performed in the temperature range of 2 K-300 K using a superconducting quantum interference device (SQUID), Cryogenic UK. Consistent with our earlier results [1], the current pellet of Co$_2$C NPs from the same batch also shows a robust *M-H* hysteresis persisting up to 300 K. At low *T*, particularly below $T_{EB}$ = 50 K, the pellet shows the EB effect. Figure 1(a) shows field cooled (FC) hysteresis loops at 2 K and 60 K, wherein a cooling field of 3 T was applied at 300 K before cooling down the sample to the desired *T*. It is evident from Fig. 1(a) that there is an asymmetry ($H_L$ (left coercive field) > $H_R$ (right coercive field)) in the *M-H* loop at 2 K, compared to the symmetric *M-H* loop of 60 K (> $T_{EB}$). Figure 1(b) shows the variation of the exchange bias field $H_{EB} = \frac{|H_L|-|H_R|}{2}$, ($H_L$ and $H_R$ are identified in Fig. 1(a)) with *T* in a log-linear plot. The blue line is the linear fit that shows the exponential *T* dependence of $H_{EB}(T)$. An earlier detailed analysis of the training effect established the presence of the EB effect in NPs of Co$_2$C [1]. AC susceptibility study in Fig. 1(c) also confirms [1] the presence of a broad peak ~ 5 K, which shifts with the excitation frequency (*f*). This peak represents the SG transition at $T_{SG}$ ~ 5 K in the Co$_2$C NPs-pellet sample[31]. Our earlier studies using $Co_2C$ NPs made from the same batch have confirmed that this peak in AC susceptibility at 5 K corresponds to an SG transition. It has also been shown that its shift in the peak with frequency obeys the Vogel-Fulcher law, which suggests an SG phase. The SG transition was also confirmed from features seen in DC magnetization studies as well (see Ref. [1]). Note that the observation of an *M-H* hysteresis (Fig. 1(a)) is not sufficient evidence for the presence of long-range magnetic order. It is known that [32, 33] in a granular magnetic system with short-range magnetic order, long-ranged dipolar interactions between the grains result in a correlated magnetic response. Such a system can exhibit *M-H* hysteresis akin to a ferromagnetic-like system, although the system intrinsically does not have any long-range magnetic order. Although we see a hysteretic *M-H* at low *T* (Fig. 1(a)), our bulk *T* -dependent susceptibility measurements reveal a disordered spin glass configuration at low $T < T_{SG}$ (Fig. 1(c)). It is therefore of interest to explore further the magnetic state in our NPs.

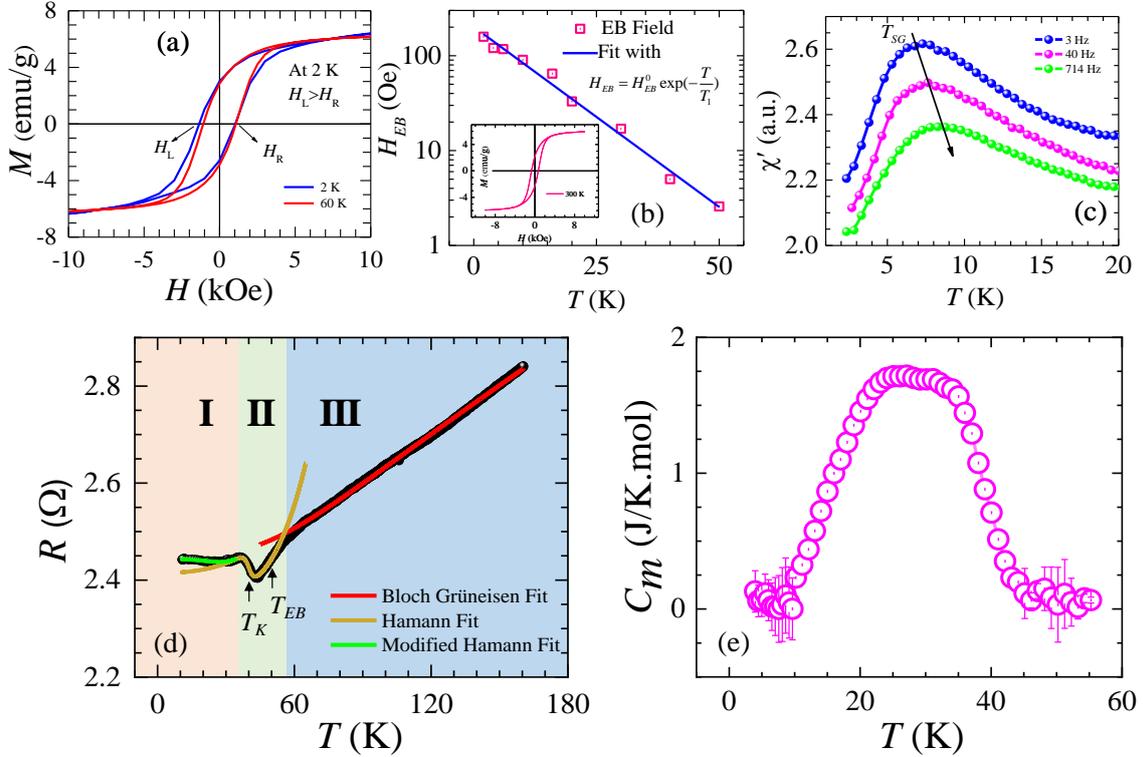

FIG. 1 (a) FC *M-H* hysteresis loops at 2 K and 60 K. (b) Exchange Bias field ($H_{EB}$) with temperature in log-linear scale. The solid line is the exponential fit presented in the log-linear scale. The inset shows the *M-H* hysteresis loop at 300 K (c) Real part ($\chi'$) of ac susceptibility with the temperature at different excitation frequencies (*f*). (d) Resistance (*R*) as a function of temperature with a constant current of 1 mA flowing through the sample. The data can be divided into three distinct regimes. The solid red curve is the fit to the data in regime III using Bloch Grüneisen function; the solid yellow curve is the fit to the data in regime II with the conventional Hamann equation with contributions from electron-electron and electron-phonon mediated terms included; in regime I, the data has been fitted with the modified Hamann equation along with the term arising from the electron-electron dominated contribution. The solid green curve gives the fit in regime I. See text for details of the fits used. (e) Magnetic contribution ($C_m$) to the specific heat with temperature.

**Electrical and thermal properties of $Co_2C$ nanoparticle pellet:**

Figure 1(d) shows the resistance, $R$ versus $T$ measurement (four probes) on the pellet of $Co_2C$ NPs. Four Gold pads were deposited on the pellet for the four-probe measurement. At 170 K the pellet has a resistivity ($\rho$) = 2 × $10^{-2}$ $\Omega$.m, and an residual resistivity ratio, RRR = $R$(170 K)/$R$(55 K) = 1.2. The relatively high resistivity of $Co_2C$ NPs-pellet compared to a typical metal suggests that the pellet does not follow the theoretical prediction of a paramagnetic metallic nature of bulk $Co_2C$ [27]. Fitting the $T$ dependence of the specific heat of the pellet ($C_{sample}$) (see Fig. 2 in SI-II) gives a Debye temperature, $\theta_D$ = 250 K. Hence as per the specific heat measurements, the $T$ regime from below 170 K until 55 K is associated with strong electron-phonon (e-ph) interactions in the system.

We note from Fig.1(d) that it is not possible to explain the $R(T)$ behavior with one single expression. Therefore the $R(T)$ data in Fig. 1(d) is analysed by dividing it into three distinct regions, viz., regimes I, II and III as seen in Fig.1(d). From the extrapolated coloured lines as shown in Fig.1(d) in the different regimes, note that there is an abrupt change in the curvature of $R$ versus $T$ dependence as one transitions across the regimes. This feature suggests different underlying physical processes become active in the magnetic NP's composite at the transition boundary between the different regimes. We see that above 55 K in region III, the data fits well to the Bloch Grüneisen function (see solid red curve) [34],

$$R(T) = R_0 + A\left(\frac{T}{\theta_D}\right)^5 \int_0^{\theta_D/T} dx . x^5 / [(e^x - 1)(1 - e^{-x})] \tag{1}$$

Here $R_0$ is the residual resistance due to defect scattering and $\theta_D = 250$ K (as determined from $C_{sample}(T)$). A good fit is obtained in region III with $R_0 = 2.45$ Ω and A = 3.52 Ω. The $R(T)$ behaviour suggests a metallic nature with strong electron-phonon scattering dominating the observed $R(T)$ behavior in this regime. Here we would like to mention that the above measurement uncovers the intrinsic metallic nature of $Co_2C$ NPs. This is confirmed as bias dependence of the local density of states (DOS) measured through Scanning Tunneling Microscopy (STM) measurements also reveals the presence of a finite DOS at the Fermi level (see Fig. 3(a) in SI-II), which is characteristic of metallic nature of $Co_2C$ NPs. We have also measured two other pellets made from the $Co_2C$ NPs, and we observe similar $R(T)$ behaviour, thereby confirming that the observed $R(T)$ behaviour is not sample specific. Note that at this juncture, we cannot deduce the extent of grain boundary scattering contribution to $R(T)$. It is likely to be present, however its contribution is subdominant compared to the Bloch Grüneisen mechanism of scattering, which gives a good fit for the $R(T)$ behavior over a wide $T$ window in regime III. The grain boundary scattering however may result in a low RRR (~ 1.2) for our samples.

Below 55 K, which is in the vicinity of $T_{EB}$ (= 50 K), the resistance drops sharply to a minimum $R_{min} = 2.4$ Ω at ~ 43 K; it increases again to a maximum followed by a secondary minimum before ultimately saturating. The appearance of a minimum in $R$ suggests two possibilities, (i) opening of a bulk gap in the DOS or (ii) electrons getting localized in the material. We first consider the possibility that a minimum feature represents Kondo localization of electrons [35-37]. The Kondo localization is associated with the localization of electrons which screen local magnetic moments present in $Co_2C$ NPs-pellet. The sharp fall below $T_{EB}$ and the consequent resistance upturn at lower temperatures is fit using two forms. The solid yellow curve in Fig. 1(d) is a fit to the data in regime II using the equation [38]

$$R(T) = R_0 + qT^2 + pT^5 + R_{K0}\left\{1 - \frac{\ln\left(\frac{T}{T_K}\right)}{\sqrt{\ln^2\left(\frac{T}{T_K}\right)+s(s+1)\pi^2}}\right\} \qquad (2)$$

The first term $R_0$ here is the constant resistance; the second term arises from contributions due to the usual electron-electron (el-el $\propto T^2$ term) interactions; the third term comes from electron-phonon (el-ph $\propto T^5$ term) interactions, and the last term is the conventional Hamann function corresponding to Kondo localization, where $T_K$ is the Kondo localization temperature. The fit gives $R_0 = 2.33\ \Omega$, $q = 1.92\times10^{-5}\ \Omega.K^{-2}$, $p = 1.95\times10^{-10}\ \Omega.K^{-5}$, $R_{K0} = 0.04\ \Omega$, Kondo temperature $T_K = 40.13$ K and average spin of magnetic site, $s = 3.5\times10^{-4}$. By comparing the $p$ and $q$ values in regime II, we see that the contributions from the el-ph interactions are far weaker than the el-el interactions. The resistance upturn can be attributed to the Hamann equation; the competition between this logarithmic increase and the $T^2$ varying decrement due to the el-el contributions results in the Kondo minimum at ~ 43 K. It may be noted that a low value of $s$ is obtained when the Hamann function is used for fitting data for $T < T_K$ [39, 40]. Although the fit captures well the features of the sharp fall and the resistivity upturn above and below the Kondo minimum, respectively, it deviates from the $R(T)$ data below ~ 35 K. The solid green curve is the fit to the data below 35 K in regime I with the equation

$$R(T) = R_0 + qT^2 + R_{K0}\left\{1 - \frac{\ln\left(\frac{\sqrt{T^2+T_W^2}}{T_K}\right)}{\sqrt{\ln^2\left(\frac{\sqrt{T^2+T_W^2}}{T_K}\right)+s(s+1)\pi^2}}\right\} \qquad (3)$$

It uses $R_0$ as the constant resistance and includes the usual el-el ($\propto T^2$ term) scattering term giving $R_0 = 2.13\ \Omega$, $q = 1.2\times10^{-4}\ \Omega\ K^{-2}$. The last term in the expression is the modified Hamann function [39-42] related to scattering from the Kondo effect and includes the additional contribution associated with Ruderman-Kittel-Kasuya-Yosida (RKKY) interaction. The strength of the RKKY interactions is $k_B T_W$, and an effective temperature replaces the $T$ in the Hamann expression $T_{eff} = \sqrt{T^2 + T_W^2}$. The above expression gives a good fit to the data in regime I with $R_{K0} = 0.19\ \Omega$, $T_W = 24.71$ K and $T_K = 40.13$ K, and $s = 0.03$, where we obtain using the modified Hamann function a more realistic estimate of $s$. Note that we also measured the $R$-$T$ of two other samples (see Figs. 3(b) and 3(c) in SI-II), which exhibit similar behaviour and have resistivity minima at the same temperature. We would like to mention that we have not found a single functional form that gives a continuous fit spanning the three different temperature regimes. Our analysis shows that the strength of the relative contribution of the different scattering terms (values of the fit parameters) gets modified across the different $T$ regimes (see Fig. 1(d)), due to the 'switching on' of different effects which modify scattering as the different $T$ boundaries of different regimes are crossed. In regime III at $T > T_{EB}$ the electron – phonon scattering determines the behavior of $R(T)$. However as one enters in regime II, viz., $T < T_{EB}$, with the onset of EB phenomenon there is significant modification to scattering effects and we observe features related to Kondo localization phenomena associated with the strengthening of electron-electron correlation effects, and all of this completely modifies the scattering features upon entering regime II. We also see this clearly demonstrated in regime II, where the dominance of electron-electron scattering effects over electron-phonon scattering effects shows up as the parameter $p \ll q$. With further decrease in $T$ there is strengthening of the electron – electron correlation effects and this results in the emergence of Kondo localization and effects of RKKY correlation contribute to modified scattering at low $T$ in regime I. Therefore in a complex system like $Co_2C$ we see the 'switching on' of different phenomena as the different $T$ regimes are crossed, which in turn significantly modify the nature of scattering events experienced by the electrons in the system. Consequently, the coefficient of the $R(T)$ fit gets modified across the different regimes. In Fig. 1(e), we plot magnetic contribution to the specific heat $C_m(T) = C_{sample}(T) - [C_{Debye}(T) + \gamma T]$, where $C_{Debye}(T) + \gamma T$ is the Debye (phononic) and electronic contribution to specific heat determined by fitting the measured specific heat behaviour of the sample ($C_{sample}(T)$) (see SI-II for details). Concomitant with the evidence of the Kondo effect seen in the electrical transport, the specific heat peak in $C_m(T)$ below $T_K$ (Fig. 1(e)), further supports the Kondo localization scenario in $Co_2C$ NPs-pellet. It is worth noting that $T_K$ is quite close to $T_{EB} = 50$ K, where the EB effect sets in. The appearance of the Kondo effect also suggests that $Co_2C$ is a strongly correlated electronic system. From the fit to $C_{sample}(T)$ (see SI-II), we obtain a high value of Sommerfeld's constant, $\gamma = 74$ mJ/K$^2$mol (note $\gamma \propto g(\varepsilon)$, where $g(\varepsilon)$ is the DOS at energy $\varepsilon$). The enhanced $\gamma$ value [43-46] suggests the presence of a strong correlation in $Co_2C$ NPs. Our earlier Scanning Tunneling Microscopy (STM) based differential conductance measurements $\frac{dI}{dV} (\propto g(\varepsilon))$ in $Co_2C$ NPs- pellet showed a strongly enhanced DOS $g(\varepsilon)$, near the Fermi level, $\varepsilon_F$ (see Fig. 3(a) in SI-II and also Fig. 4(a) inset in ref.[1]; there appears a peak near STM bias voltage = 0 V), which also suggests strong correlations present in $Co_2C$. DFT calculations on $Co_2C$ reported in our earlier studies [1] do not suggest any bulk gap opening in the system. We also do not see evidence of any gap-like feature in the STM spectra around $\varepsilon_F$ (see Fig. 3(a) in SI-II). Therefore, opening up of a bulk gap in the DOS is unlikely to be responsible for a minimum in $R(T)$ near $T_K$. Thus, the above studies show that strong correlation effects in $Co_2C$ NPs is responsible for the onset of Kondo localization at $T_K$. The Kondo localization appears just below the onset of the EB effect at $T_{EB}$. Therefore, the nature of resistivity as inferred by the bulk transport measurement is in good agreement with the independent estimations of the local

STM-based differential conductance and the specific heat measurements and therefore establishes the intrinsic material property.

**Exploring the magnetic state of Co$_2$C NP pellet at different *T* using μSR studies:**

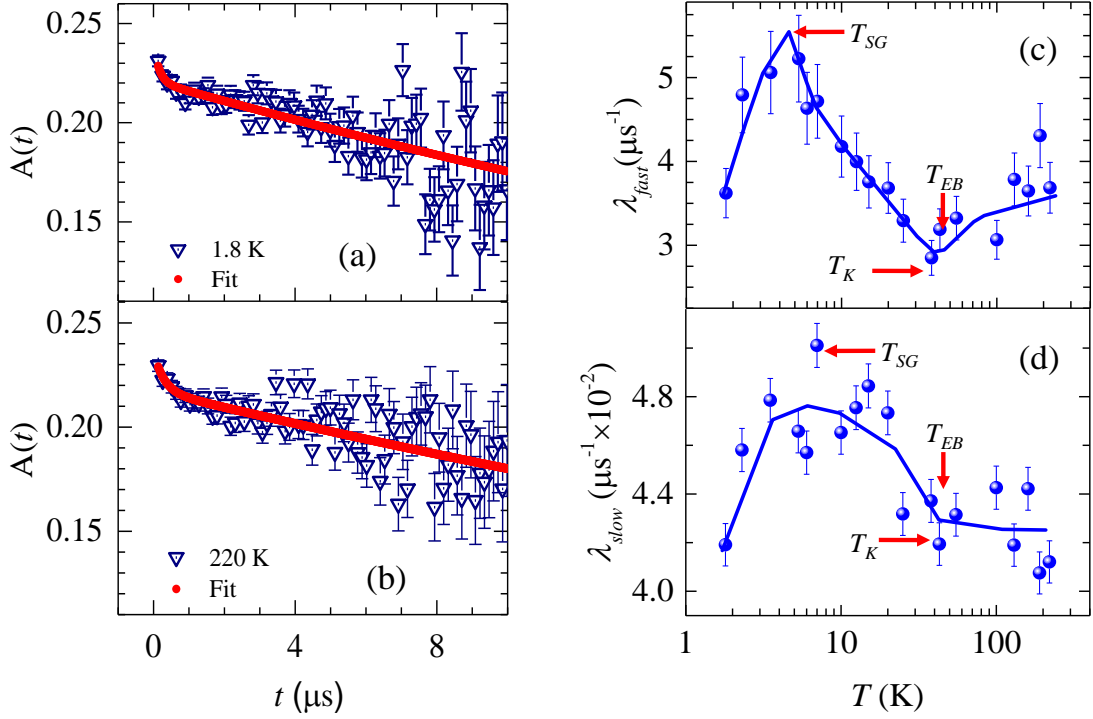

FIG. 2 (a) and (b) show a representative ZF-μSR time spectra at 1.8 K and 220 K, respectively. The solid lines in the spectra are fit to Eqn. 5. (c) & (d) show the behaviour of relaxation rate of fast & slow components with temperature.

We study the magnetic configuration in the Co$_2$C NPs- pellet using μSR measurements. The μSR measurements were carried out in the temperature range of 1.8 K-220 K using the HiFi spectrometer at ISIS pulsed muon facility, Rutherford Appleton Laboratory, UK [47]. The Co$_2$C NPs-pellet was loaded onto the standard Ag (99.99%) sample holder used for μSR due to its time-independent background signal. In our μSR measurements, fully spin-polarized muons are implanted into the sample, which starts precessing around the local magnetic field present at its stopping site in the magnetically ordered state. The decay of muons to positrons is recorded with time using forward (*F*) and backward (*B*) detectors placed with respect to muons' initial polarization direction. The measured physical quantity, the evolution of muon spin polarization viz., asymmetry, is defined as follows:

$$A(t) = \frac{N_B(t) - \alpha N_F(t)}{N_B(t) + \alpha N_F(t)} \quad (4)$$

$N_B(T)$ and $N_F(T)$ are the number of positron counts in backward and forward detectors at time *t*, respectively, *α* is a relative efficiency factor of the two detectors. The value of *α* (=1.2870) was estimated from the small transverse field (TF = 20 G) measurements at high temperatures. Figures 2(a) and (b) show the raw zero field (ZF)-μSR time series *A*(*t*) spectra at 1.8 K and 220 K, respectively. The presence of any long-range magnetic order in the sample will produce oscillations in the zero field (ZF) μSR asymmetry spectrum for a smaller ordered moment but will show a loss of 2/3 of the initial muon asymmetry for a large ordered moment. We observe no oscillatory signal in the ZF-μSR-*A*(*t*) spectra and no loss of the initial asymmetry, which suggests the absence of any long-range magnetic ordering present in the Co$_2$C NPs-pellet down to 1.8 K [48-50]. We find that the ZF-μSR-*A*(t) spectra can be fitted only by using two different relaxation rates. It appears that the sample has one component which relaxes rapidly within the first ~1 μs and another component that relaxes slowly (above ~1 μs). The ZF-μSR spectra for all *T* [50-54] are fitted with the following expression,

$$A(t) = A_{BG} + A_{fast} \exp(-\lambda_{fast} t) + A_{slow} \exp(-\lambda_{slow} t) \quad (5)$$

Where $\lambda_{fast}$ and $\lambda_{slow}$ are fast and slow components of muon spin relaxation rates. $A_{BG}$ is a constant background, and $A_{fast}$, and $A_{slow}$ are the corresponding initial asymmetry values, with $A_{fast} \neq A_{slow}$ and $\lambda_{fast} \neq \lambda_{slow}$. The behaviour of $\lambda_{fast}$ and $\lambda_{slow}$ is shown in Figs. 2(c), 2(d) in log scale. We see that $\lambda_{fast}$ is nearly two orders of magnitude larger than $\lambda_{slow}$. The $\lambda_{slow}$ is however an order of magnitude higher than the background relaxation rate ($\lambda$) [55-57] of muons embedded in high purity paramagnetic Ag sample holder. Also note that non monotonic changes in the $T$ dependence of $\lambda_{fast}$ and $\lambda_{slow}$ are not characteristic of a paramagnetic state in the $Co_2C$ NP pellet. The two distinct relaxation, $\lambda_{fast}$ and $\lambda_{slow}$ rates suggest the presence of two inequivalent muon stopping sites in the NPs target, where the muon experiences different strengths of local magnetic field environment [58, 59].

Muons rapidly precess in the strong local field leading to rapid decay in the ZF asymmetry spectra. Therefore the $\lambda_{fast}$ for $t < 1$ μs corresponds to muons embedded in a relatively strong local magnetic field environment in the sample. In contrast, muons that are embedded in an environment with a disordered magnetic configuration, they experience a fluctuating magnetic field with a small average local field. Due to this, the precessional frequency of these muons is low and consequently ZF – asymmetry relaxes slowly with a rate $\lambda_{slow}$. The amplitudes, $A_{fast}$ and $A_{slow}$ are proportional to the sample volume fractions for the two muon stopping sites [60]. ZF-μSR spectra are not very sensitive to the variation in $A_{slow}$ and $A_{fast}$ as a function of temperature. We find them to be almost temperature independent, therefore in all subsequent analyses, we will use the $T$-independent average value of $A_{slow}$ and $A_{fast}$. Using the total initial $A(t)$, $A_{slow}$ and $A_{fast}$ values, we calculate the fractions of the average muon population at the two sites. Using this we determine that the fast-relaxing rate is associated with the magnetic fraction having short-range order (~ 14% of the sample volume), and the slower relaxation rate is associated with the disordered magnetic fraction (~ 86% of the sample volume). It is clear that there is a local magnetic moment distribution inside the sample and the fluctuating magnetic moments do not exhibit onset of long range order down to the $T$ we have studied.

In Fig. 2(c) and 2(d), below 50 K (~ $T_{EB}$, $T_K$), the $\lambda_{fast}$ rapidly increases and reaches a maximum near 5 K which is very close to the SG peak in $\chi'(T)$ at $T_{SG}$ ~ 6 K (cf. Fig. 1(c)). For $T < T_{SG}$ viz., in the SG state, the $\lambda_{fast}(T)$ begins to decrease. Due to EB effect, below $T_{EB}$ the fluctuating local magnetic fields in the sample are stabilized and hence the average local field increases. Therefore $\lambda_{fast}$ shows a pronounced increase from below $T_{EB}$. Note that the $\lambda_{fast}(T)$ component is sensitive to modifications in the dynamics of these fluctuating local fields, as it shows a cusp-like feature at a $T$ corresponding to the onset of the spin glass phase, viz., at $T = T_{SG}$ (~ 6 K) (see Fig. 2(c)). It is also worth mentioning here that a similar cusp-like feature in $\lambda_{fast}(T)$ has been found near the SG transition in other materials as well [49, 51, 61]. Somewhat similar features as $\lambda_{fast}(T)$ are also shown by $\lambda_{slow}(T)$ values (solid blue line) as well. From the behaviour of $\lambda_{fast}(T)$ and $\lambda_{slow}(T)$ in Fig. 2(c) and 2(d), it appears that in the regime between $T_{EB}$ and $T_{SG}$, the magnetic environment of the muons is modified compared to that above $T_{EB}$.

We explore this fluctuating magnetic state in our $Co_2C$ sample further by measuring the local magnetic field distribution at different $T$'s using transverse field (TF)-μSR spectra. The TF-μSR measurements were carried out with the applied field of 20 G at different $T$. In TF-μSR geometry, the magnetic field is applied perpendicular to muon's initial spin polarization. Figure 3(a) shows the measured TF-μSR asymmetry $A(t)$ data captured at 2.3 K (solid black line). The Fast Fourier Transform (FFT) of the measured TF μSR-$A(t)$ spectra at 2.3 K is shown in Fig. 3(b). The sharp peak in the FFT amplitude at 0.27 MHz corresponds to the externally applied transverse field of 19.92 G. We note that beyond the main peak corresponding to the applied field in Fig. 3(b), there is a broad tail with significant FFT amplitude, which extends up to higher frequencies of about 5 MHz.

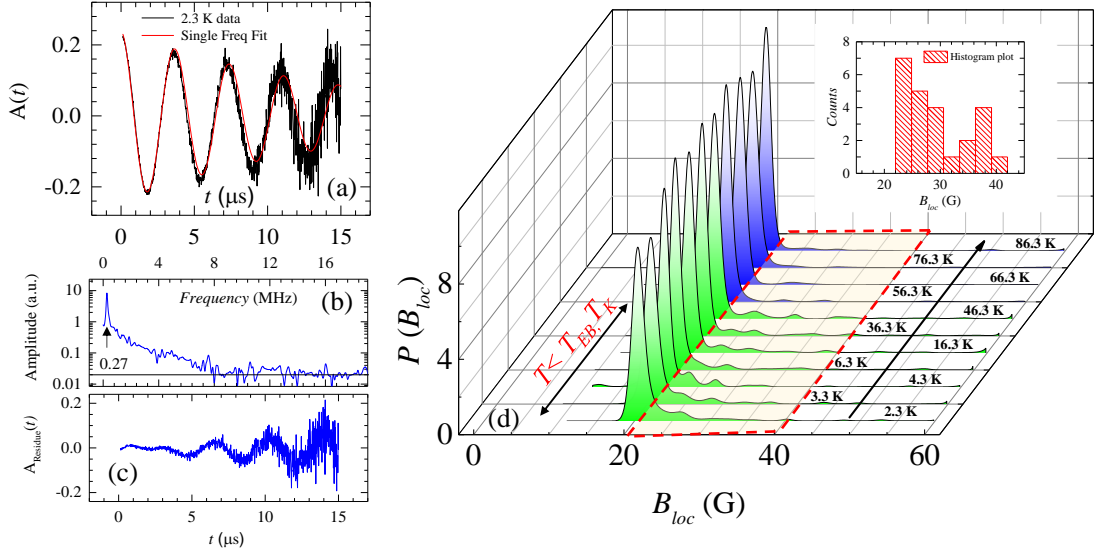

FIG. 3 (a) TF-µSR spectra at 2.3 K measured in 20 G, the red curve is the fitted data using single frequency (0.27 MHz) in muon decay function, i.e., $F(t) = A exp(-\lambda_T t) cos(\omega t + \phi)$, where $\omega = \gamma_\mu B_{loc}$, $\phi$ is the phase factor, and $B_{loc}$ is the local magnetic field experienced by muons. (b) Fast Fourier Transform of TF- $A(t)$ spectra at 2.3 K shown in Fig. 3(a). (c) The residual asymmetry obtained after fitting of 2.3 K data. (d) Field distribution of the local field in $Co_2C$ NPs at different temperatures, obtained by maximum entropy transformation of the TF-µSR spectra. The inset shows the histogram plot of the local field values (obtained by counting the number of times of different $B_{loc}$ values observed in the $P(B_{loc})$ peaks below 50 K) inside the sample.

The higher frequencies in the tail correspond to the presence of the local field higher than the applied field inside the sample. We attempt to fit the measured $A(t)$ data in Fig. 3(a) with a single frequency ($\omega = \gamma_\mu B_{loc}$) function of the form $F(t) = A \exp(-\lambda_T t) \cos(\omega t + \phi)$, $\phi$ is the phase factor. We use $\omega = \gamma_\mu B_{loc} = 0.27$ MHz, obtained from FFT, where $\gamma_\mu = 2\pi \times 135.5342$ MHz/T (in supplementary section III Fig. 4, we plot the residue obtained with a direct fit with $F(t)$ where $\omega$ (a fitting parameter) gives a best fit value of 0.275MHz). The best fit $F(t)$ curve to $A(t)$ is shown in Fig. 3(a), the fit yields values of $A = 0.242$, $\lambda_T = 0.068$ µs$^{-1}$, and $\phi = 0.077$ rad. In the above expression for $F(t)$, the term $A exp(-\lambda_T t) = A_{fast} e^{-\lambda_{fast} t} + A_{slow} e^{-\lambda_{slow} t}$. Note that in this expression, over longer time scales, $\lambda_T \sim \lambda_{slow}$, the fast relaxation component decays out within $t \leq 0.5$ µs (which is far less than even one full period in the TF spectra). In supplementary section III we show that the by fitting TF data using the approximate, $F(t) = A\exp(-\lambda_T t)\cos(\omega t + \phi)$ form the $\lambda_T$ is comparable to the $\lambda_{slow}$ determined from the earlier ZF measurements. In Fig. 3(c), we plot $A_{residue}(t) = A(t) - F(t)$, which shows the presence of significant residual asymmetry using a single frequency (0.275 MHz) in the expression for $F(t)$, to fit the $A(t)$ data. This suggests the insufficiency of using a single frequency, $f = 0.275$ MHz for fitting our data, which inturn suggests including additional frequency components to analyse the TF asymmetry spectra. Note that similar alternative fits used for analysing typical TF-μSR spectra of the type, $F(t) = A\,exp(-\lambda_T t)\exp(-\frac{\sigma^2 t^2}{2})\cos(\omega t + \phi)$, yields very low values of σ~ $0.005\mu s^{-1}$ and we find it does not yield any significantly better fit compared to that in Fig.3(a). Using the modified $F(t)$, oscillations still persist in the $A_{residue}(t)$, similar to that seen in Fig.3(c). The need for incorporating higher frequencies in our analysis of the asymmetry spectra, is further supported by noting the presence of an extended frequency tail above 0.27 MHz in the FFT spectra of Fig. 3(b), indicating additional frequency components contribute to the signal. For this purpose, we use the maximum entropy spectral analysis [62, 63] which captures the presence of higher $f$ components in the TF-μSR-$A(t)$. Since multiple $f$'s corresponds to the presence of a distribution of local internal fields, $B_{loc}$, the analysis gives the distribution $P(B_{loc})$ experienced by the muons (where $\omega = 2\pi f = \gamma_\mu B_{loc}$). Figure 3(d) main panel shows $P(B_{loc})$ versus $B_{loc}$ measured at different $T$.

In Fig. 3 (d) note that the peak in $P(B_{loc})$ is at 20 G, the applied transverse magnetic field. While at higher $T$, only the 20 G peak is visible in the $P(B_{loc})$ plot; however, as $T$ is lowered, additional satellite peaks appear in the $P(B_{loc})$ distribution below 50 K (= $T_{EB}$). The peaks in $P(B_{loc})$ above 20 G are more pronounced below $T = 43$ K, which is close to $T_K$ (see the evolution of peaks above 20 G in $P(B_{loc})$ plot inside the red dashed square regions in Fig. 3(d)). From the $P(B_{loc})$ at different $T$'s, we count the number of times peaks in $B_{loc}$ (>20 G) appear in $P(B_{loc})$. The histogram of the counts versus $B_{loc}$ is shown in the inset of Fig. 3(d). The distribution of $B_{loc}$ (> 20 G) in the inset of Fig. 3(d) spreads out between ~ 23 G to 40 G. Note that at ISIS the applied TF was set to 20 G with an accuracy of ~ ±1.2 G. The spread in the applied field is much lower than the field distribution seen in Fig.3(d) inset. Thus, we clearly see that, at $T < 50$ K $= T_{EB}$, a magnetic state with a broad distribution of local magnetic fields appears in the $Co_2C$ NPs-pellet.

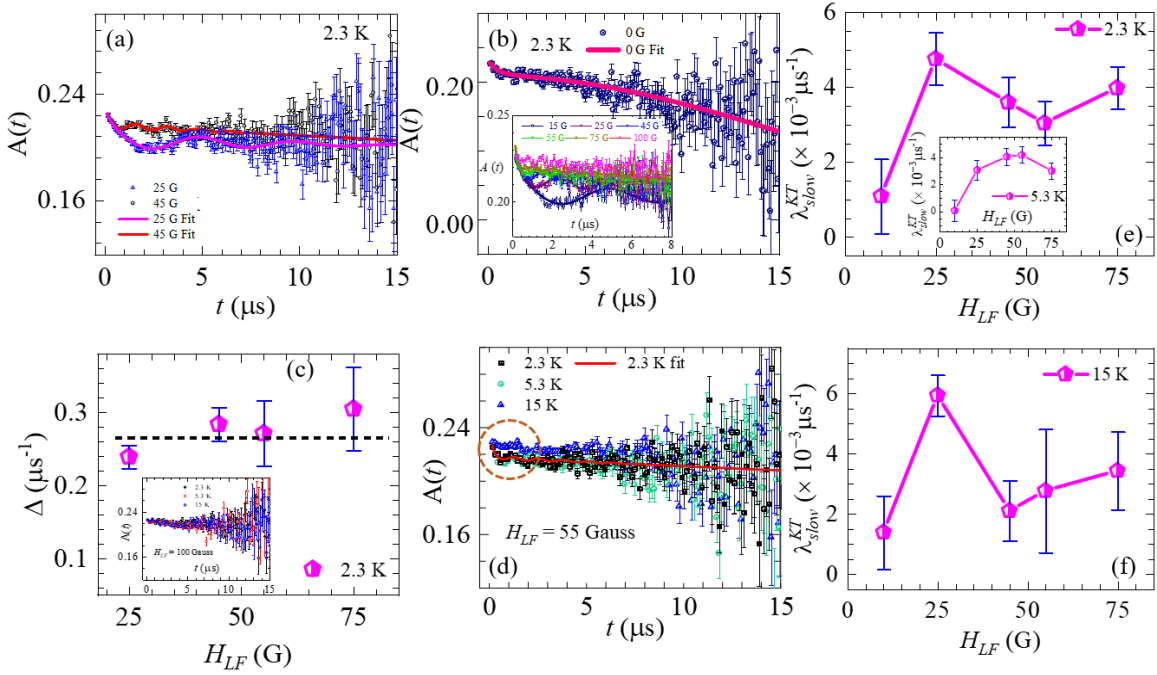

FIG. 4 (a) Longitudinal field (LF) $\mu$SR measurements performed at two different fields of 25 G and 45 G and at 2.3 K. The solid lines are the fits to the data using the static Kubo-Toyabe form (Eqn. 6.) (b) shows the ZF $\mu$SR-$A(t)$ spectra; the solid line is the fit with Eqn. 6. Inset shows the evolution of the spectra under different LF, viz., 15, 25, 45, 55, 75, 100 G and at 2.3 K. (c) Variation of the width of the local field distribution ($\Delta$) with field ($H_{LF}$), obtained at 2.3 K when fit with Eqn. 6. In the inset is shown the $\mu$SR-$A(t)$ spectra at different temperatures under $H_{LF} = 100$ G. Note that irrespective of temperature, the fast relaxation component is absent (see text for details). (d) shows the longitudinal field (LF) $\mu$SR-$A(t)$ spectra taken at 2.3 K, 5.3 K, and 15 K for $H_{LF} = 55$ G and the fit to 2.3 K data with Eqn. 6 with the solid red line. See that at 2.3K and 5.3 K, the fast and slow relaxing components are clearly seen (see the encircled region) while at 15 K, the fast relaxation is absent. (e) shows the behaviour of relaxation rate of the slow components with $H_{LF}$ at 2.3 K; in the inset is shown relaxation rate of the slow components with $H_{LF}$ at 5.3 K . (f) shows the behaviour of relaxation rate of the slow components with $H_{LF}$ at 15 K;

We now investigate the dynamics of the magnetic fluctuations present in our $Co_2C$-NP system using longitudinal field (LF) μSR measurements at 2.3 K (< $T_{SG} = 6$ K). In LF μSR measurements, the longitudinal field ($H_{LF}$) is applied in the direction of muon's spin polarization, the implanted muons precess around the resultant magnetic field, which is the net of $H_{LF}$ and the internal local field at the muon site. For two different applied longitudinal fields, $H_{LF}$ = 25 G and 45 G, Fig. 4(a) shows the LF-μSR asymmetry time-series spectra, $A(t)$, captured at 2.3 K. The progressive loss of decay in $A(t)$ signal with increasing $H_{LF}$ in the inset of Fig.4(b), suggests the muons get strongly polarized along the applied field direction. At large enough fields the muon spins are completely polarised by the applied field and they decouple from the environment as no relaxation in the asymmetry signal is observed [48, 64]. To analyse dynamics of spins the LF-μSR asymmetry spectra, we use a form based on the well-known Kubo-Toyabe function $G_Z(H_{LF},\Delta,t)$ in the ZF-μSR-$A(t)$ fitting in Eqn. (5) [59, 60] viz.,

$$A(t) = A_{fast} \exp(-\lambda_{fast}^{KT} t) + A_{slow} \exp(-\lambda_{slow}^{KT} t) \times G_Z(H_{LF}, \Delta, t) \qquad (6)$$

Where $G_Z(H_{LF},\Delta,t)$ is the static Kubo-Toyabe function (see supplementary SI-III), $\Delta/\gamma_\mu$ is the width of the distribution of $B_{loc}$ at the muon sites. Note from Fig. 4(a) that, Eqn. (6) gives a good fit for both 25 G and 45 G LF-spectra's at 2.3 K and Fig. 4(b) shows Eqn. (6) also fits the ZF-μSR spectra at 2.3 K by setting $H_{LF} = 0$ G in the equation. We find that by fitting the ZF-μSR spectra obtained at different $T$ using Eqn. (6), yields $\lambda_{fast}^{KT}(T)$, whose magnitude and $T$ behavior are similar to $\lambda_{fast}(T)$ obtained earlier in Fig. 2(c) (see SI section III Fig. 5). Note that the $\lambda_{slow}^{KT}(T)$, which characterizes the long time dynamics of the muons in the local magnetic field environment, is approximately three orders of magnitude smaller than $\lambda_{fast}^{KT}$. The $T$-independent $A_{fast}$ and $A_{slow}$ values are determined from the ZF-μSR spectra. Figure 4(c) shows the variation of $\Delta(H_{LF})$ at 2.3 K. The nearly constant $\Delta$ (~ 0.3 μs$^{-1}$) with $H_{LF}$ suggests the average width of the distribution of the local field seen by the muon remains almost unchanged in the SG regime with varying $H_{LF}$. Inset of Fig. 4(c) shows that at 100 G the polarized muons have completely decoupled from their local magnetic environment at all $T$ and do not show any decay. The $\Delta \sim 0.3$ μs$^{-1}$ corresponds to an average field spread of $\sim 3.5$ G. Note that this field spread is much smaller than the spread in $B_{loc}$ determined in inset of Fig. 3(d) through TF measurements. Figure 4(d) shows that $H_{LF}$ = 55 Oe is not sufficient to fully polarize the muons and decouple them from their magnetic environment at 2.3 K. The spectra in Fig. 4(d) shows that without changing $H_{LF}$, increasing $T$ to 15 K leads to the decoupling of muons. The above suggests that at low $T$ there are low-frequency fluctuating magnetic fields arising from magnetically ordered regions in the pellet (possible regions with short range order) with which the muons are coupled.

Figures 4(e) and 4(f) show the dynamics of magnetic fluctuations. By comparing the $\lambda_{slow}^{KT}$ behaviour above 20 G in Figs. 4(e) and 4(f) we see a drop of $\sim 25\%$ at 15 K compared to the values at 2.3 K or 5.3 K (for $H_{LF}$ > 20 G). At 15 K the slow relaxation (low $\lambda_{slow}^{KT}$ value) in the presence of a relatively high $H_{LF}$ (> 20 G) suggests that muons have decoupled from the local magnetic field environment and they only sense the fluctuating background nuclear field. In a similar $H_{LF}$ range (> 20 G), from 2.3 to 10 K, the muons are coupled to the strong local field environment which is not overcome by the applied $H_{LF}$ and the relaxation rate $\lambda_{slow}^{KT}$ is comparatively higher which suggests that the fluctuations in the system are low compared to its high temperature (15 K) counterpart [65]. In this low $T$ regime where the Co$_2$C NP sample host significant local field, further lowering the $H_{LF}$ from 20 G results in a significant drop in $\lambda_{slow}^{KT}$ (Fig. 4(e)). As the field increases, there is a subsequent slowing down of the dynamical fluctuations in this low temperature regime. The sustained nature of dynamical fluctuations at low $T$ has been seen in f-electron systems and spin glass systems [51, 57, 61, 66-68]. Below the glass freezing temperature where the moments are frozen into a glassy configurations, a significant drop in the muSR relaxation rate has been observed in the past [51, 57]. These results are consistent with our susceptibility studies showing the presence of glassy like phase below 6 K. It is interesting to note that the temperature and field dependence of $\lambda_{fast}^{KT}$ also exhibits a comparable behavior to that of $\lambda_{slow}^{KT}$ (see supplementary information Section IV Fig. 6 ) due to freezing of spins dynamics as the sample enters the glassy regime at low temperatures ($< T_{SG}$).

Our study of the ZF-μSR spectra of the fast and slow amplitude in Fig. 2 shows that Co$_2$C NPs-pellet has about 14% short-range ordered fraction, and 86% is a magnetically disordered fraction. Our rough estimation from fitting the virgin *M-H* curve at 2 K (see SI-IV for details of the fit), shows that the typical size of the cluster (island) with short-range magnetic order is ~ 15 $\mathring{A}$ which is much smaller than a single nanoparticle size of ~ 40 nm. Our μSR studies have already shown that ~ 14% of the pellet volume has these islands with short-range order distributed randomly in a sea of disordered spin configurations. The relaxation rate associated with the short-range ordered fraction $\lambda_{fast}(T)$ shows a significant increase between $T_{EB}$ and $T_{SG}$ (Fig. 2(c)). This increase corresponds to the rapid increase in the depolarization rate of the muons, which suggests more interaction between the muons with the local internal fields stabilizing in the sample. Hence the increase in $\lambda_{fast}(T)$ between $T_{EB}$ and $T_{SG}$ corresponds to the short-range ordered magnetic state becoming more stable in the NPs. We believe the increase in $\lambda_{fast}(T)$ and $\lambda_{slow}(T)$ below 50 K signifies the onset of exchange interaction ($J_{EB} = K_B T_{EB}$) at the interface between the fraction with short-range order and magnetically disordered fractions. Thus, at $T_{EB}$, a magnetic interface forms with the $J_{EB}$ stabilizing at the interface. This $J_{EB}$ couples the magnetically short range-ordered and disordered fractions on two sides of the interface and which in turn helps in stabilizing the local internal fields in the system. As the fluctuations of the random moments decrease below $T_{EB}$, the strong correlations in Co$_2$C lead to Kondo screening of the stabilized moments at random sites by the conduction electrons below $T_K$ (Fig. 1(d)). At $T > T_{EB}$, the $R(T)$ behaviour is governed by electron-phonon scattering and there is no significant magnetic contribution to it. As $T$ is lowered below $T_{EB}$, strengthening electron-electron correlation effects result in the formation of the Kondo localized states below $T_K$. This leads to enhanced scattering and $R(T)$ increase. So below $T_K$ the 14 % regions in the pellet with short range magnetic order begin to make their presence felt. At lower $T \ll T_K$, viz., in the RKKY regime, the remaining 86% regions also contribute to the $R(T)$ behavior. Our analysis in Fig. 1(d) shows a crossover from the Kondo-screened scenario to an RKKY interaction-dominated regime. In this regime, the islands with short-range order interact via the RKKY interaction mediated by the conduction electrons. At low $T$ and low fields, the magnetic clusters are coupled by RKKY interactions. Frustration in the interactions possible leads to a frozen spin glass like configuration with very slow dynamics at these low $T$'s. The $\lambda_{fast}$ and $\lambda_{slow}$ significantly decreasing below $T_{SG}$ (Fig. 2(c)) is due to entry into a frozen SG regime, whose evidence is also seen in the LF-muSR results. At $T < T_{EB}$ with the onset of $J_{EB}$ at the interface, the fluctuations of the moments around the interface are suppressed, and the internal local fields are stabilized. As seen in Fig. 3(d), this shows up as a stable broad local field distribution ($P(B_{loc})$) below $T_K$ and $T_{EB}$. Hence, we propose that in the $Co_2C$ NPs-pellet, a magnetic interface forms below $T_{EB}$. This interface separates a disordered fraction and a short-range ordered fraction with an exchange interaction that couples the two fractions. The formation of the magnetic interface with $J_{EB}$ below $T_{EB}$ could be triggered by the onset of some additional weak AFM interactions in the system. Earlier DFT work [23] showed that the $Co_2C$ NPs have a structurally ordered core with bulk-like atomic coordination of Co; however, there is a significant increase in the Co atoms with lower coordination near the surface of the NPs. The structurally distorted shell of the NPs may harbour weak AFM interaction whose effect becomes prominent below $T_{EB}$. This AFM interaction competing with FM interaction in the $Co_2C$ system leads to frustration and transition into an SG state at low $T < T_{SG}$. Note that our NPs-pellet is an agglomeration of grains. Hence there is a complex meandering manifold of grain boundaries crisscrossing our sample. The grain boundary has no preferred orientation. Thus, the EB effect which arises at such a meandering interface is also unlikely to exhibit any preferred orientation.

In summary, the Co$_2$C NP pellet system is a complex and strongly interacting system whose presence is revealed through effects that switch on below different characteristic temperature scales. Our ZF $\mu$SR and TF $\mu$SR measurements reveal that the mean-field magnetic landscape of the Co$_2$C NP pellet can be described as clusters or islands with short-range magnetic order. These islands are embedded in a sea of magnetically disordered medium. The clusters and the surrounding medium have a complex set of interactions coupling them. The first coupling relates to exchange bias interaction which switches on below $T_{EB}$ = 50 K. With lowering of $T$ below $T_{EB}$ there is are strong electronic correlations effects in Co$_2$C NP pellet resulting in Kondo Localization to set in below $T_K \sim$ 40.13 K. The occurrence of Kondo localization in the $T_K$ regime is found to be near $T_{EB}$. At lower $T$ ($< T_K$) there are features of another interaction, viz., the RKKY interactions which produce a frozen spin-glass phase at $T < T_{SG}$. Further detailed theoretical and experimental investigations into the complex magnetic character of Co$_2$C NP pellet.

**Acknowledgment:** S. S. Banerjee would like to acknowledge the funding support from the Department of Science and Technology (DST), the Government of India and RAL UK program, and IIT Kanpur, DST-SERB SUPRA program. SSB would also like to thank the ISIS Facility for beam time on the HiFi spectrometer, RB1968044. Suprotim Saha acknowledges the Prime Minister's Research Fellows (PMRF) Scheme of the Ministry of Human Resource Development, Govt. of India, for funding support.

# Probing the strongly correlated magnetic state of Co$_2$C nanoparticles at low temperatures using μSR

Nirmal Roy[1], P C Mahato[1], Suprotim Saha[1], M. Telling[2], J. S. Lord[2], D T Adroja[2,3], S. S. Banerjee[1,+]

[1]*Indian Institute of Technology Kanpur, Kanpur, Uttar Pradesh 208016, India*

[2]*ISIS Facility, Rutherford Appleton Laboratory, Chilton, Didcot Oxon OX11 0QX, United Kingdom*

[3]*Highly Correlated Matter Research Group, Physics Department, University of Johannesburg, P.O. Box 524, Auckland Park 2006, South Africa*

**Supplementary Information Section I:**

**Sample Preparation:**

Samples are prepared using the one-pot polyol reduction process [1]. First, we dissolved 1.02 g of NaOH in 20 ml tetraethylene glycol (TEG) in a container by heating it to 100 °C. Another 30 ml of TEG is taken into a 100 ml European flask containing 2.5 mmol of Co (CH$_3$CO$_2$)$_2$·4 H$_2$O and 0.75 g of polyvinylpyrrolidone (PVP). A magnetic stirrer is then used to mix the mixture for 20 minutes at room temperature. The solution is then poured into the European flask, and the mixture is heated to 373 K for 30 minutes to eliminate the water. The solution is heated further to the boiling point of TEG (583 K) for 1 h. The solution is allowed to cool spontaneously to room temperature. The solution is then diluted with ethanol, and the precipitated nanoparticles (NPs) are extracted from the solution with a rare earth magnet. The leftover solution is drained, and the precipitate is washed with ethanol many times. The extracted NPs are dried in a vacuum oven, and the powders are compacted into pellets. The pellets do not have any preferred orientation because they are made up of NPs of size of ~ 40 nm.

**Sample Characterization:**

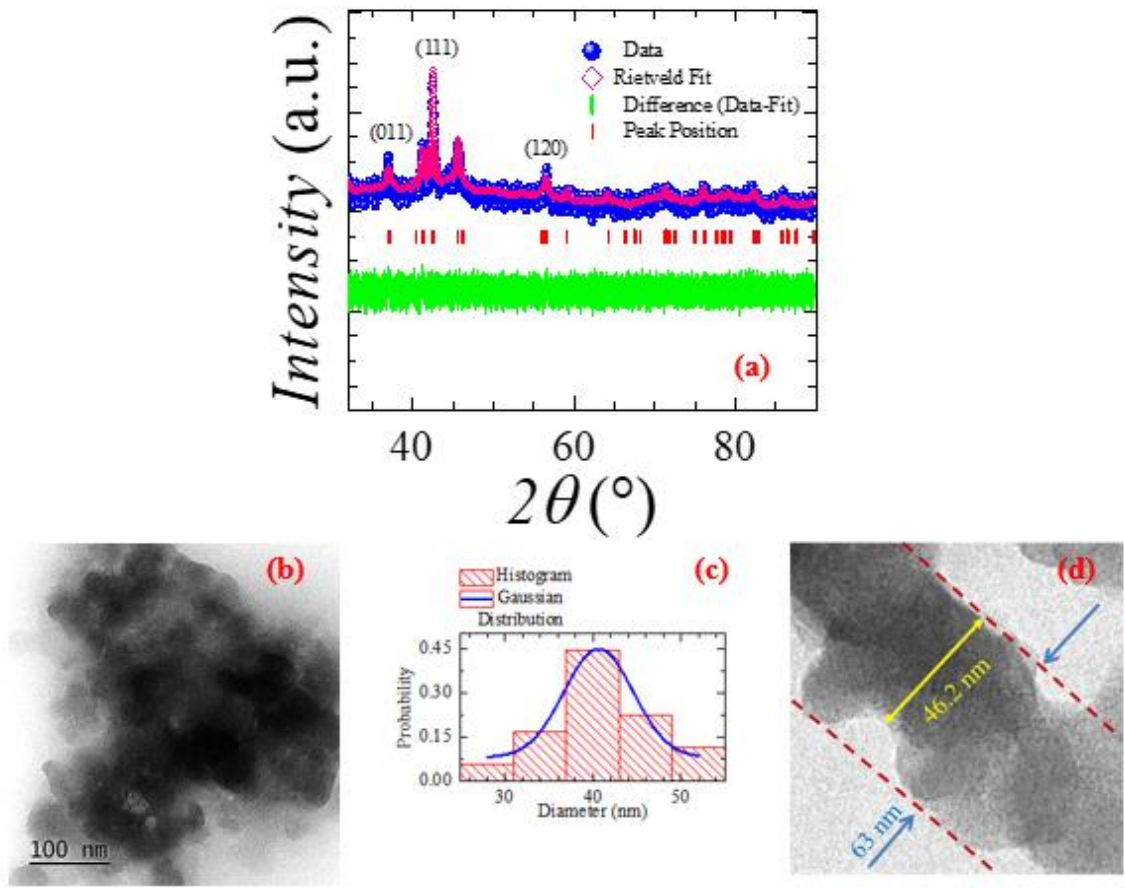

Figure 1. (a) X-ray powder diffraction pattern of Co$_2$C–NPs. (b)TEM image of the synthesised nanoparticles. (c) Histogram plot of particle size distribution and Gaussian fit to the distribution giving average particle diameter of (40 ±7.8) nm where the standard deviation of Gaussian fit is 7.8 nm. (d) TEM image of a line of agglomerated NPs.

X-ray diffractions (XRD) are performed to determine the phase and crystal structure. XRD analysis using Rietveld refinement method are shown in Fig. 1(a) confirming the presence of pure Co$_2$C phase with orthorhombic crystal structure (see Ref. [1] for details). X-ray studies of our sample show that the X-ray diffraction (XRD) peaks are sharp (see Fig. 1(a)). We use the XRD peak width to estimate the average particle size of the NPs using the Debye Scherrer equation $D = \frac{\xi\lambda}{\beta \cos\theta}$, where $D$ is the average crystallite size. The parameter $\xi$ is the dimensionless shape factor ∼ 0.9, $\lambda$ (= 0.15406 nm for Cu K$\alpha$) is the X-ray wavelength, $\beta$ is the full width at half maximum of XRD peak, and $\theta$ is the Bragg angle. Using the value of $2\theta = 42.5°$ and $\beta = 0.28°$, we estimate D = (30.7 ± 7) nm. Some of these detailed morphological characterizations of the as-synthesised Co$_2$C NPs were already presented in our earlier paper [1]. The TEM image of synthesised nanoparticles and histogram plot of particle size are shown in Fig. 1(b) and (c) to determine the average particle size. The average particle size is (40 ±7.8) nm, where 7.8 nm is the standard deviation (SD) of Gaussian fit. Because of their magnetic nature, individual Co$_2$C nanoparticles aggregate, making it impossible to get images of individual nanoparticles. A single line of agglomerated particles is depicted in Fig. 1(d) above. The typical diameter of the nanoparticles is of the order (40 ±7.8) nm, as shown in Fig. 1(d).

**Supplementary Information Section II:**

**Heat Capacity Measurements:**

We measured the temperature dependence of specific heat of the Co$_2$C nanoparticles (NPs) and the background from the sample holder and the system without the sample. First, we measure the heat capacity of the background without the sample. For our background measurements, on the sample holder, instead of the sample, we place the amount of apiezon grease necessary to glue our sample to the holder. The specific heat of the sample ($C_{sample}$), shown in the Fig. 2(a) below, is obtained by subtracting the background specific heat from the total specific heat measured with the sample. $C_{sample}(T)$ is analyzed to determine the magnetic contribution to specific heat. The total specific heat of the sample is

$$C_{sample} = C_{Debye}(T) + \gamma T + C_m(T) \qquad (1)$$

where $C_{Debye}(T) = \dfrac{9nR}{\left(\theta_D/T\right)^3} \int_0^{\theta_D/T} \left(\dfrac{x^4 \exp(x)}{(\exp(x)-1)^2}\right) dx$, $n$ and $R$ are number density in a single formula unit of the material and molar gas constant, $\theta_D$ is Debye temperature, and $C_m(T)$ is the magnetic contribution to specific heat while γ is the Sommerfeld's constant. We break up the specific heat data into two temperature regimes, namely, regime (i) below 20 K, where the specific heat behaviour is governed essentially by the electronic contribution, and (ii) above 50 K, where the phononic contribution dominates. Based on the fitting parameters obtained in these regimes, we show the plot of the function $C_{Debye}(T) + \gamma T$ in Fig. 2(a) below. From the fit, we have obtained a $\gamma \sim (74 \pm 10)$ mJ/K$^2$.mole and $\theta_D = 250$ K. In Fig. 2(b) below, we determine the magnetic contribution to specific heat using $C_m(T) = C_{sample} - [C_{Debye}(T) + \gamma T]$. It is clear that there is a significant increase in the magnetic contribution to specific heat, which occurs below 40 K, corresponding to the onset of the Kondo effect in the NPs.

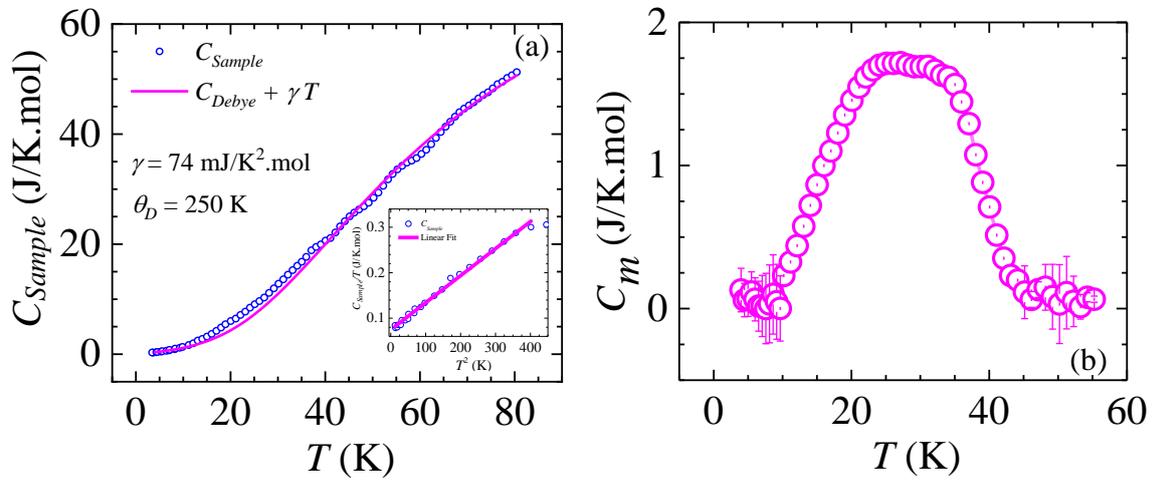

Figure 2. (a) shows temperature dependence of specific heat of the Co$_2$C NPs fitted with $C_{Debye}(T) + \gamma T$. $\gamma$ has been determined from the linear fit of the specific heat data at $T < 21$ K ;see inset. (b) shows the magnetic contribution to the specific heat.

**Scanning Tunneling Microscopy (STM) Measurements:**

STM measurements are performed using a Quazar Technologies room temperature STM (NanoRev. 4.0). The average of 15 measurements of differential conductance $dI/dV$ ($\propto$ density of state (DOS)) is performed as a function of bias voltage $V$ (-0.8 to 0.8 V) at 300 K, and $I$ is the tunneling current. The Figure 3(a) below shows enhanced $dI/dV$ ($\propto$ DOS) at Fermi energy $\varepsilon_F$ ($V = 0$). The DOS shows absence of any gap at the Fermi Level and the metallic nature of $Co_2C$ NPs. The error bars represent the extent of scatter in the measured local DOS, obtained from doing measurements at different locations on the pellet.

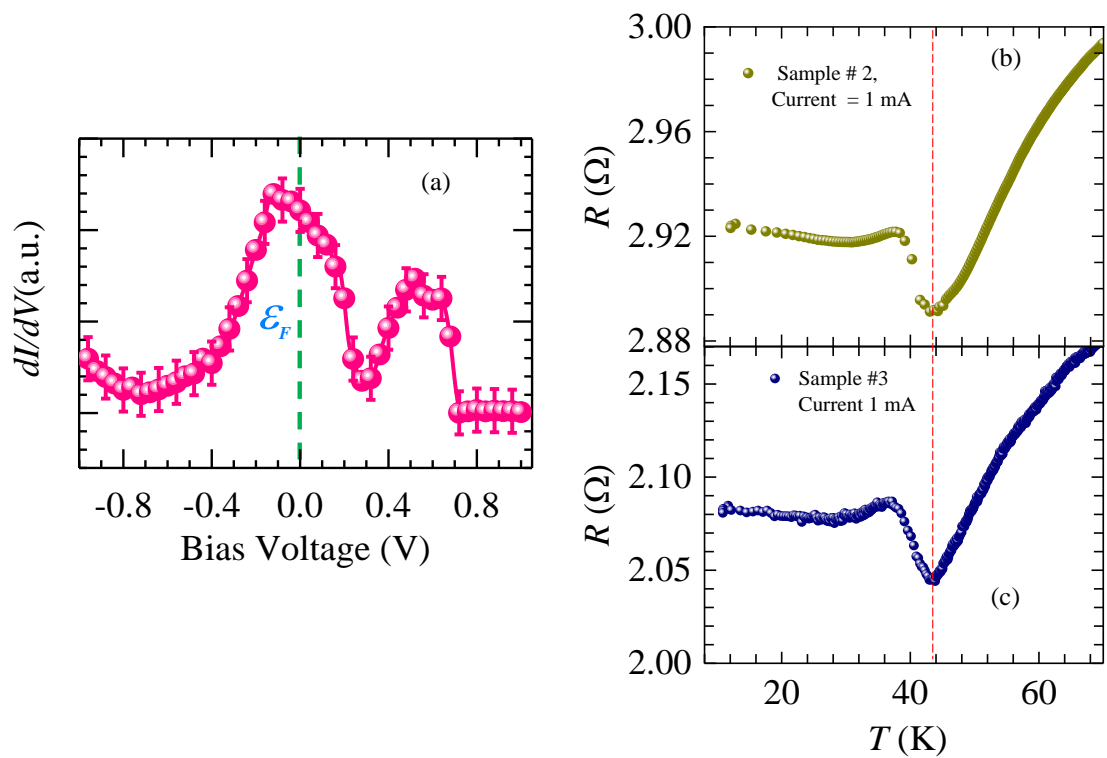

Figure 3. (a) shows the average of 15 measurements of differential conductance d$I$/d$V$ ($\propto$ DOS) measured on $Co_2C$ NPs of size 40 nm as a function of bias voltage $V$ at 300 K, and $I$ is the tunneling current. (b) and (c) shows $R$-$T$ of samples #2 and #3 of $Co_2C$ NPs-pellets. Both pellets show similar behavior with a resistivity minimum at the same temperature (red dashed line).

We also measured the *R-T* of two other samples (see Figs. 3(b) and (c)), which exhibit similar behavior and have resistivity minima at the same temperature. This confirms that the features seen in the resistivity measurement are not sample-dependent.

**Supplementary Information Section III:**

**Single frequency fit of Transverse field (TF) data:**

The TF asymmetry spectra at 2.3 K is fitted with the function $F(t) = A\, exp(-\lambda_T t) \cos(\omega t + \phi)$ and we keep $\omega$ as a free fitting parameter. The fit gives a value of $\omega$ = 0.275 MHz, which is close to the FFT peak at 0.27 MHz (see Fig, 3(b) in manuscript). Despite allowing the frequency to vary in the fit, it still shows a considerable residual asymmetry (see Fig. 4 below).

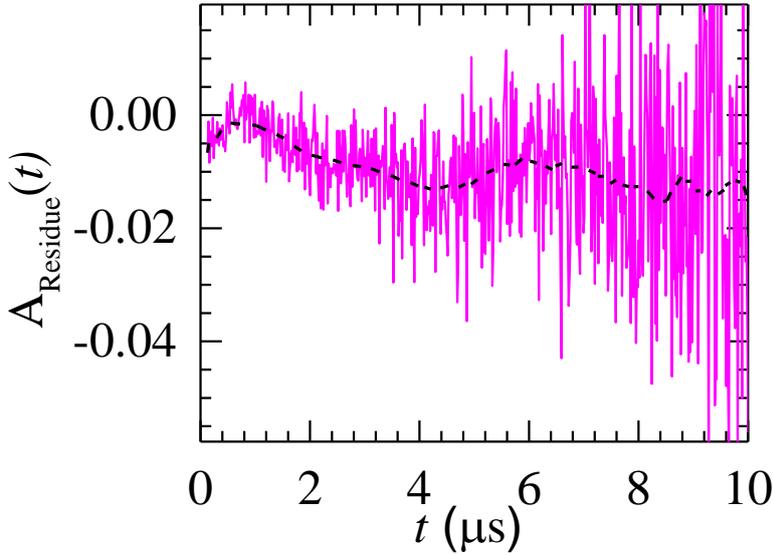

Figure 4: shows the residual asymmetry of fit. The black dashed curve represents the modulating trend of the 50 point adjacent average value of the residue. The modulation in the average value of residue shows the inadequacy of a single frequency fit for the transverse asymmetry data.

**Comparison of slow relaxing component obtained from ZF and TF spectra**.

In our main manuscript (see Fig. 3(a)) we had used a model $F(t) = A\, exp(-\lambda_T t) \cos(\omega t + \phi)$ to fit the TF asymmetry data. It may be noted that the term $A\, exp(-\lambda_T t) = A_{fast}\, exp(-\lambda_{fast} t) + A_{slow}\, exp(-\lambda_{slow} t)$. Note that the dominant contribution $A\exp(-\lambda_T t)$ will be from the slow relaxation term as the fast relaxation component decays out within $t \leq 0.5$ μs. The residue plot in the Fig. 3(c) of our manuscript had suggested that *F(t)* fit to TF data in Fig. 3(a) is not the best fit, as there are additional frequency components which are present in the data. We have described this further in detail in our MS. we have evaluated the value of $\lambda_T$ at different *T* using the above *F(t)* and compared them with the $\lambda_{slow}$ values obtained from ZF data in the table below. To within order of magnitude we find a fairly close agreement between the $\lambda_T$ and $\lambda_{slow}$ values despite the limited validity of the *F(t)* function to describe the TF data.

| Temperature (K) | $\lambda_T$ from F(t) fit from TF data ($\mu s^{-1}$) | $\lambda_{slow}$ from ZF data ($\mu s^{-1}$) |
|---|---|---|
| 2.3 | 0.057 ± 0.001 | 0.0458± 0.0009 |
| 5.3 | 0.042 ±0.001 | 0.0466± 0.0009 |
| 13 | 0.0472±0.001 | 0.0476± 0.0009 |
| 15 | 0.0475±0.001 | 0.0484± 0.0009 |

**Zero Field (ZF) and Longitudinal Field Kubo Toyabe (KT) Function**

In zero-field condition, the static Kubo Toyabe (KT) function is given by

$$G_Z(t) = A\left[\frac{1}{3} + \frac{2}{3}(1 - \Delta^2 t^2)e^{(-(1/2)\Delta^2 t^2)}\right]$$

where *A* is the muon asymmetry spectra at time *t* = 0 and $\Delta$ is the local field distribution. In the presence of a longitudinal field $\mu_0 H_{LF} = \frac{\omega_0}{\gamma_\mu}$, where $\omega_0$ is the muon's Larmor frequency corresponding to the magnetic field that it experiences and $\gamma_\mu$ is the muon's gyromagnetic ratio, the static Kubo Toyabe function gets modified as[2, 3];

$$G_Z(t) = A\left[1 - \frac{2\Delta^2}{\omega_0^2}\left(1 - \cos(\omega_0 t)e^{-(1/2)\Delta^2 t^2}\right) + 2\frac{\Delta^4}{\omega_0^4}\omega_0\int_0^\tau \sin(\omega_0 \tau)e^{-(1/2)\Delta^2 t^2}d\tau\right]$$

**Fitting ZF μSR Data with Static Kubo Toyabe (KT) Function :**

We fit the ZF μSR *A(t)* spectra with the combination of a fast-relaxing component and slow relaxing component multiplied by Zero Field Kubo Toyabe (KT) function as given below

$$A(t) = A_{\text{fast}}e^{-\lambda_{fast}^{KT}t} + A_{\text{slow}}e^{\lambda_{slow}^{KT}t}\{1/3 + 2/3(1-\Delta^2 t^2)\exp(-1/2\Delta^2 t^2)\}$$

Where $A_{\text{fast}}$, $\lambda_{fast}^{KT}$, $A_{\text{slow}}$, $\lambda_{slow}^{KT}$ are the asymmetry and relaxation rate of the fast and slow components, respectively; the term within curly bracket is the zero-field static KT function where delta is the spread of the local field distribution. During fitting we have fixed $A_{\text{fast}}$, $A_{\text{slow}}$ at their high temperature values as they do not seem to vary appreciably with temperature. Here, we show the variation of $\lambda_{fast}^{KT}(T)$; it shows a dip around $T_{\text{EB}}$ and $T_{\text{K}}$ and then rises sharply to a hump around $T_{\text{SG}}$. Therefore, $\lambda_{fast}^{KT}$ captures well the characteristic temperature scales of the system.

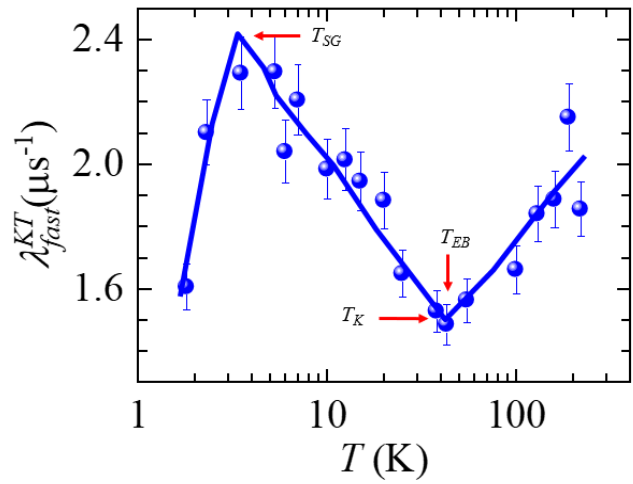

Figure 5. Variation of $\lambda_{fast}^{KT}$ as a function of temperature

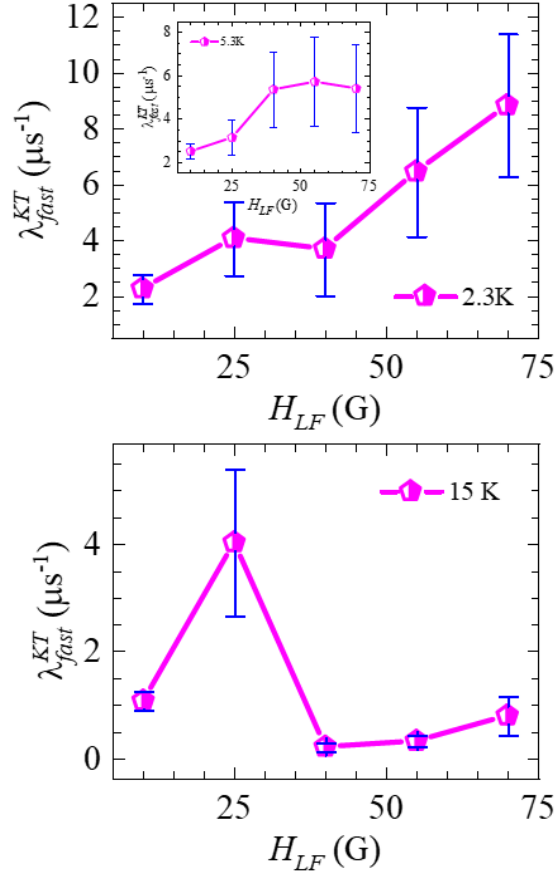

Figure 6. (a) Variation of $\lambda_{fast}^{KT}$ with $H_{LF}$ at 2.3 K; inset shows the $\lambda_{fast}^{KT}$ with $H_{LF}$ at 5.3 K. (b) Variation of $\lambda_{fast}^{KT}$ with $H_{LF}$ at 15 K

Above shows the relaxation rate of the fast component in the LF-µSR spectra, viz., $\lambda_{fast}^{KT}$ (see eqn. 6) in the main MS. As $T$ changes from 2.3 K to 15 K, Figs. 6(a) and (b), show that at higher fields e.g., at 55 Oe, there is a significant drop in relaxation rate [4] from about 7 µs$^{-1}$ to less than 1 µs$^{-1}$. At higher $T$ (15 K) and higher $H_{LF}$, the muons decouple from the local magnetic field environment, and the relaxation rate is associated with the muons sensing the background nuclear fields only. At low $T$ of 2.3 K and 5.3 K we had seen at low $H_{LF}$ the muons do not decouple from the local magnetic field environment and there they sense the presence of significant local moments in the muons environment which is different from the background nuclear fields. Furthermore, in this low $T$ regime by going to low $H_{LF}$ value, we see in Fig.6(a) that there is a drop in the relaxation rate, $\lambda_{fast}^{KT}$, as muons sense the frozen glassy magnetic state which sets in the Co$_2$C NP sample (cf. discussion from the main MS on the glassy state seen below $T_{SG}$ in Fig.1).

**Supplementary Information Section IV:**

We estimate below the typical size of the regions with short-range order. In order to do this, we determine the average magnetic moment of the cluster regions with short-range order. For doing this, instead of fitting the virgin $M(H)$ curve with a simple Langevin function ($L(\frac{\mu H}{k_B T})$) which implicitly assumes a uniform distribution of magnetic moments ($\mu$), we use the modified Langevin function [5]. The modified Langevin form considers variance in the values of magnetic moments ($\mu$) in the nanocomposite, explicitly a log-normal distribution $f(\mu) = \frac{1}{\mu s\sqrt{2\pi}} \exp(-[\ln(\mu/\xi)]^2/2s^2)$, where $s$ is the variance and $\xi$ is related to the average magnetic moment through, $\mu_{avg} = \xi\sqrt{e^{s^2}}$. The $M(H)$ is fit to the modified Langevin function (as shown in Figure 5 below), viz., $M(H) = N_{cluster} \int_0^\infty \mu f(\mu) L(\frac{\mu H}{k_B T}) d\mu$, where $N_{Cluster}$ is a number density parameter. Using $N_{Cluster}$, $s$ and $\xi$ as the fitting parameters, we fit the virgin $M(H)$ curve at 2 K using the modified Langevin function; we get $\mu_{avg} = 20\ \mu_B$ per cluster of the Co$_2$C NPs-pellet with $N_{cluster} = 5 \times 10^{19}$ per gram. With magnetic moment per cobalt (Co) atom = 0.066 $\mu_B$ at 2 K in Co$_2$C [1], and there are 6 Co atoms per Co$_2$C unit cell, one estimates an average size of 50 unit cells per cluster of the region with short-range magnetic order. As each Co$_2$C unit cell has a volume of 56.65 $Å^3$, and assuming a spherical cluster (island) with short-range magnetic order, the typical cluster (island) size is ~ 15 Å which is much smaller than a single nanoparticle size of (~ 40 nm).

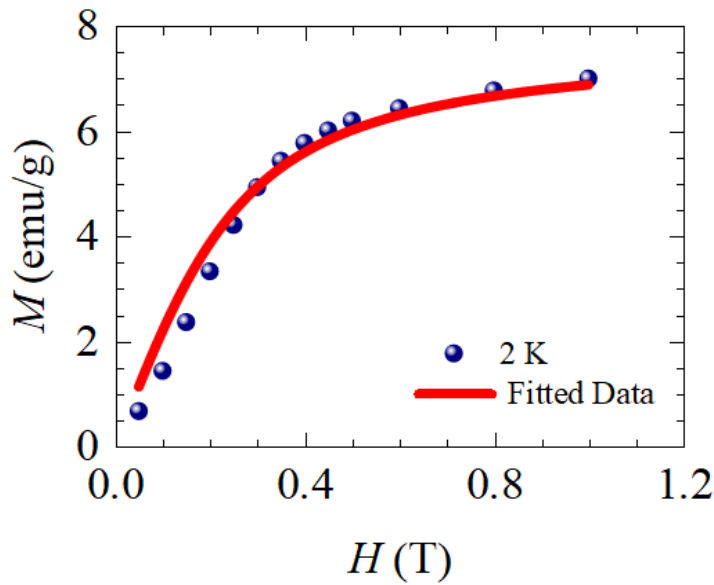

Figure 7. Magnetization versus applied field (*M-H*) curve at 2 K. The solid red line is the fit to the data with the modified Langevin equation (see text for details).